\documentclass{article}

\usepackage{own}
\usepackage{prooftree}

\newcommand{\sas}{\,\&\,}

\title{A Rule-Based Logic for Quantum Information}
\author{Olivier Brunet \\ Leibniz - IMAG - University of Grenoble \\ 46, avenue Félix Viallet - 38031 GRENOBLE Cedex \\ France}

\begin{document}

\maketitle

\section*{Introduction}

The logic modelled by orthomodular lattices, usually referred to (in a rather controversial way) as quantum logic, constitutes the main formalism for the logical study of quantum mechanics. First introduced in the thirties by von Neumann and Birkhoff \cite{Birkhoff36QuantumLogic}, the logic modelled by orthomodular lattices (usually referred to, in a rather controversial way, as quantum logic) has since then been used as the starting point of most attempts to understand the quantum world from a logical and algebraical point of view \cite{Svozil98Book, Hughes89Book, DallaChiara2001QuantumLogic}.

\ 

The key difference between classical logics and quantum ones is that distributivity is replaced by a much weaker law, call orthomodularity:

$$ a \leq b \ \Rightarrow \,\ a = b \wedge (a \vee b^\bot)$$

In particular, this particularity implies that the conjunction becomes rather difficult to handle. To quote Jean-Yves Girard, ``there is a fine negation (the orthogonal complement), but nothing like a decent conjunction'' \cite{GirardQuantic}.

\ 

A convenient way to formalize a logic for a proof-theoretical approach is to use tableaux or sequent calculi. To this respect, there have been some attempts to provide such formulations of quantum logic, but most of them only deal with a weaker logic, called minimal quantum logic (one can refer to \cite{Cutland82QLSeq, Nishi94MQL, Egly99Tableaux} and also to \cite{DallaChiara2001QuantumLogic}).

\ 

In the present article, we explore a new approach for the study of orthomodular lattices, where we replace the problematic conjunction by a binary operator, called the Sasaki projection. We present a characterization of orthomodular lattices based on the use of an algebraic version of the Sasaki projection operator (together with orthocomplementation) rather than on the conjunction. We then define of a new logic, which we call Sasaki Orthologic, which is closely related to quantum logic, and provide a rule-based definition of this logic.

\section{A Framework for Classical Partial Knowledge}

As it is well-known (see \cite{Ptak91Book}), to every physical system, one can associate an orthoposet whose elements correspond to properties of the system which can be verified experimentally. In that case, the ordering of this structure corresponds to the entailment between properties and the orthocomplementation corresponds to the negation of a property.

\ 

In \cite{Brunet03JLC, Brunet04IJTP}, we introduce a general algebraic framework, called {\em representation system}, to model a notion of point of view of a system together with the way information can flow from one point of view to another. In this approach, it is possible to study knowledge about a system with explicit references to points of view, and also to reason about it by considering the existence of these points of view and the way they relate to each other. 

An important restriction to this formalism corresponds to the case where the system is observed in a classical way, so that the verifiable properties at a given moment form a finite (or more generally complete) boolean algebra and where we consider that the boolean algebra is actually a boolean subalgebra of the orthoposet associated to the system.

In particular, as we will show in the following, it is possible to study and characterize some properties of the orthoposet associated to a system by considering its boolean subalgebras, that is by considering the different ways one can observe it classically.

\  

We first introduce the notion of compatibility which relates two elements corresponding to properties which can be verified simultaneously in a classical way. Formally, this means that there is a boolean subalgebra which contains both elements belong.

\begin{Def}[Compatibility]
Two elements $x,y \in \mc P$ are said to be {\em compatible} if and only if there exists a boolean subalgebra $\mc B \subseteq \mc P$ such that $\coll{x,y} \subseteq \mc B$.
\end{Def}

As shown in \cite{Brunet04IQSA}, it is possible to characterize orthomodular posets and orthomodular lattices simply by imposing some conditions about the existence of particular boolean subalgebras, of particular points of view of the system. We first provide a characterization of orthomodular posets.

\begin{Prop} \label{Prop:OMP}
Let $\mc P$ be an orthoposet such that:
$$ \fall {x,y \in \mc P} x \leq y \Rightarrow x \,C\, y $$
Then $\mc P$ is an orthomodular poset.
\end{Prop}
\begin{Proof}
Given two elements such that $x \leq y$, they are compatible so that there exists a boolean subalgebra of $\mc P$ which contains them both. As a consequence, the join $x \vee y^\bot$ exists, and moreover, one has $ x = y \wedge (x \vee y^\bot)$ since all these calculations take place in a boolean algebra.
\end{Proof}

Intuitively, this means that if comparable elements can be verified simultaneously (from a single point of view), then the considered orthoposet is actually an orthomodular poset. By demanding an additional condition on the existence of particular boolean subalgebras, it is even possible to ensure that the structure is an orthomodular lattice (refer to \cite{Brunet04IQSA} for a detailled presentation). This condition can be justified as follows: given two elements $x$ and $y$, if they are not compatible, one can however expect to have an element $z$ which is compatible with $x$, which is greater than $y$ (it corresponds to a more general property), but is the least such element.
Formally, this condition can be expressed as follows:

\begin{Prop}\label{Prop:OML}
Let $\mc P$ be an orthoposet such that:
\begin{gather}
\fall {x,y \in \mc P} x \leq y \Rightarrow x \,C\, y \\
\fall {x,y \in \mc P} \fexist {z \in \mc P} \left\{\begin{array}{l} y \leq z \\ \text{and \ } x \,C\, z \\ \text{and \ } \fall {t \in \mc P} \paren {y \leq t \cand x \,C\, t} \Rightarrow z \leq t \end{array}\right. \label{Eq:Pixy}
\end{gather}
Then $\mc P$ is an orthomodular lattice.
\end{Prop}
\begin{Proof}
One only needs to show that $\mc P$ is a lattice, since from proposition \ref{Prop:OMP}, we already know that it is an orthomodular poset. For this, let $\pi_x(y)$ denote the element $z$ as defined in equation \ref{Eq:Pixy}. From its definition, it directly follows that $y \leq \pi_x(y)$ and if $y \leq z$, then $\pi_x(y) \leq \pi_x(z)$. Now, let us show that:
\begin{equation} \fall {x,y,z} \pi_x(y) \wedge x \leq z \ \Rightarrow \,\ \pi_x(z^\bot) \wedge x \leq y^\bot \label{Eq:PiGalois} \end{equation}
If $\pi_x(y) \wedge x \leq z$, then $z^\bot \leq (\pi_x(y))^\bot \vee x^\bot$, so that $\pi_x(z^\bot) \leq \pi_x \paren{(\pi_x(y))^\bot \vee x^\bot}$. But since $(\pi_x(y))^\bot \vee x^\bot$ is compatible with $x$, this implies that $\pi_x(z^\bot) \leq (\pi_x(y))^\bot \vee x^\bot$. As a consequence, one has:
$$ \pi_x(z^\bot) \wedge x \leq \paren{(\pi_x(y))^\bot \vee x^\bot} \wedge x \leq (\pi_x(y))^\bot \wedge x \leq (\pi_x(y))^\bot \leq y^\bot$$

Now, let $x$ and $y$ be two elements of $\mc P$, and define $z$ as $\pi_y \paren {(\pi_y(x^\bot) \wedge y)^\bot} \wedge y$. We will show that $z$ is the meet of $x$ and $y$. One obviously has $z \leq y$. Now, since $\pi_y(x^\bot) \wedge x \leq \pi_y(x^\bot) \wedge x$, it follows that directly that $z \leq x$ using equation \ref{Eq:PiGalois}. Suppose now that an element $t \in \mc P$ verifies $t \leq x$ and $t \leq y$. Since $t \leq y$, they are compatible so that computation involving these two elements can be done as in a boolean algebra. Now, from $t \leq x$, one has $x^\bot \leq t^\bot$ and then $\pi_y(x^\bot) \leq t^\bot$ (one has $t^\bot = \pi_y(t^\bot)$). From this, it follows that $(t^\bot \wedge y)^\bot \leq (\pi_y(x^\bot) \wedge y)^\bot$ and finally that $t = (t^\bot \wedge y)^\bot \wedge y \leq \pi_y \paren {(\pi_y(x^\bot) \wedge y)^\bot} \wedge y$.

Thus, we have shown that for every $x,y \in \mc P$, their meet is defined and equals $\pi_y \paren {(\pi_y(x^\bot) \wedge y)^\bot} \wedge y$.
\end{Proof}

This characterization of orthomodular lattices stresses the importance of elements of the form $\pi_x(y)$ and $\pi_x(y) \wedge x$. They can be expressed easily in terms of usual ortholattice operations: 
\begin{Prop}
Given an orthomodular lattice $\mc P$, for all $x,y \in \mc P$, one has $\pi_x(y) \wedge x = (y \vee x^\bot) \wedge x$ and $\pi_x(y) = (y \vee x) \wedge (y \vee x^\bot)$.
\end{Prop}
This way, one can recognize that $\pi_x(y) \wedge x$ corresponds to the operation usually called {\em Sasaki projection} and proposition \ref{Prop:OML} show that this operation has a central role in the study of orthomodular lattices.

\ 

In the following, we introduce an abstract version of this operation, and show the relationship between orthoposets equipped with this operation and orthomodular lattices.

\section{Sasaki Orthoposets}

In this section, we introduce {\em Sasaki orthoposets} which are orthoposets equipped with a total binary operation $\&$. This operation corresponds to a generalization of the Sasaki projection.

\begin{Def}[Sasaki Orthoposets]
A {\em Sasaki orthoposet} is a tuple $\tuple{P,\leq,\bot,\&}$ where $\tuple{P, \leq, \bot}$ is an bounded orthoposet and such that $\& : P^2 \rightarrow P$ verifies:
\begin{align}
a \leq b \ \Rightarrow \,\ & a \sas c \leq b \sas c & \text{L-Monotony} \\
& a \sas b \leq b & \text{R-Reduction} \\
a \leq b \ \Rightarrow \,\ & a = a \sas b & \text{Orthomodularity} \\
a \sas b \leq c \ \Rightarrow \,\ & c^\bot \sas b \leq a^\bot & \text{Galois}
\end{align}
\end{Def}

As expected, an orthomodular lattice can be seen as a Sasaki orthoposet by using the original Sasaki projection (which maps the pair $\pair a b$ to $ b \wedge (b^\bot \vee a)$).

\begin{Prop}
Every orthomodular lattice $\tuple{L,\leq,\bot,\wedge}$ can be turned into a Sasaki orthoposet $\tuple{L, \leq, \bot, \&_\wedge}$ with $\&_\wedge$ defined as:
$$ a \, \&_\wedge \, b = b \wedge (b^\bot \vee a) $$
\end{Prop}
\begin{Proof}
The {\em L-Monotony} and {\em R-Reduction} properties follow directly from the fact that an orthomodular lattice is a lattice. The {\em Orthomodularity} property is in that case exactly the orthomodularity condition: if $a \leq b$ then $a = b \wedge (b^\bot \vee a)$. Finally, the {\em Galois} condition reflects the following inequality:
\begin{equation} \label{Eq:GaloisConnection}
b \wedge \paren {b^\bot \vee a} \leq c \Leftrightarrow a \leq b^\bot \vee (b \wedge c)
\end{equation}
which holds in every orthomodular lattice.
\end{Proof}

The {\em Galois} property takes its name from the fact that inequality \ref{Eq:GaloisConnection} corresponds to the Galois connection $\pair {\triangleright_b}{\triangleleft_b}$ there $\triangleright_b \, a = b \wedge \paren{b^\bot \vee a}$ and $\triangleleft_b \, a = b^\bot \vee (b \wedge a)$.  Further references about these structures can be found in \cite{Birkhoff67Book} or in \cite{Erne92GaloisPrimer}.

\ 

However, Sasaki orthoposets do not constitute a generalization of orthomodular lattices, as we now show that the two notions are equivalent.

\begin{Prop}
Every Sasaki orthoposet $\tuple{P,\leq,\bot,\&}$ can be turned into an orthomodular lattice $\tuple{P,\leq,\bot,\wedge_\&}$ where $\wedge_\&$ is defined as:
$$ a \wedge_\& b = \paren{a^\bot \sas b}^\bot \sas b $$
\end{Prop}
\begin{Proof}
We first need to show that $\wedge_\&$ actually corresponds to the meet operation. It is obvious from the {\em R-Reduction} property that $a \wedge_\& b \leq b$. From $a^\bot \sas b \leq a^\bot \sas b$, it follows using the {\em Galois} property that $(a^\bot \sas b)^\bot \sas b \leq a$ or equivalently that $a \wedge_\& b \leq a$.

Now, let $c$ be an element of $P$ such that $c \leq a$ and $c \leq b$. Using {\em Orthomodularity}, one has $c \sas b = c$ and so, $c \sas b \leq a$ which implies $a^\bot \sas b \leq c^\bot$ using the {\em Galois} property. As a consequence, one has $c \leq \paren{a^\bot \sas b}^\bot$. Now, considering the {\em L-Monotony} property, one has $c \sas b \leq a \wedge_\& b$ and finally $c \leq a \wedge_\& b$ since $c = c \sas b$ due to the {\em Orthomodularity} property.

Thus, we have shown that $\wedge_\&$ corresponds to the meet operation, so that $\tuple{P,\leq,\bot,\wedge_\&}$ is an ortholattice.

\ 

We now show that the considered structure is actually an orthomodular lattice by verifying that the orthomodular inequality holds:
$$ a \leq b \ \Rightarrow a = b \wedge_\& (b \wedge_\& a^\bot) $$
Developing the expression on the right-hand side of this equality, one gets $(((a \sas b)^\bot \sas b) \sas b)^\bot \sas b$. Let us remark that due to the {\em Orthomodularity} condition, this is equivalent to $((a \sas b)^\bot \sas b)^\bot \sas b$.

Let us first prove that if $a \leq b$, then $a \leq ((a \sas b)^\bot \sas b)^\bot \sas b$.
\begin{align*}
& a \sas b \leq a \sas b \\
\Rightarrow \ & \paren{a \sas b}^\bot \sas b \leq a^\bot & \text{Galois} \\
\Rightarrow \ & a \leq (\paren{a \sas b}^\bot \sas b)^\bot \\
\Rightarrow \ & a \sas b \leq (\paren{a \sas b}^\bot \sas b)^\bot \sas b & \text{L-Monotony} \\
\Rightarrow \ & a \leq (\paren{a \sas b}^\bot \sas b)^\bot \sas b & \text{Orthomodularity}
\end{align*}
Conversely, we prove that if $a \leq b$, then $((a \sas b)^\bot \sas b)^\bot \sas b \leq a$:
\begin{align*}
& a \sas b \leq a & \text{Orthomodularity} \\
\Rightarrow \ & a^\bot \leq (a \sas b)^\bot \\
\Rightarrow \ & a^\bot \sas b \leq (a \sas b)^\bot \sas b & \text{L-Monotony} \\
\Rightarrow \ & ((a \sas b)^\bot \sas b)^\bot \sas b \leq a & \text{Galois}
\end{align*}
\end{Proof}

These two propositions show that there is a deep similitude between the two types of structures. Actually, there is a one-to-one correspondance between them since the Sasaki hook $\&$ and the meet operation $\wedge$ are entirely determined by the partial-order relation:
\begin{Prop}
Given a Sasaki orthoposet $\tuple{P,\leq,\bot,\&}$, one has:
$$ \fall {a,b \in P} a \sas b = b \wedge_\& (b \wedge_\& a^\bot)^\bot $$
where $\wedge_\&$ is defined as above. Expressed another way, one has $\& = \&_{\wedge_\&}$.
\end{Prop}
\begin{Proof}
The two sides of this equality can be shown as follows:
\begin{align*}
& b \leq b \\
\Rightarrow \ & (a \sas b)^\bot \sas b \leq b & \text{R-Reduction} \\
\Rightarrow \ & (a \sas b)^\bot \sas b \leq ((a \sas b)^\bot \sas b) \sas b & \text{Orthomodularity} \\
\Rightarrow \ & (((a \sas b)^\bot \sas b) \sas b)^\bot \sas b \leq a \sas b & \text{Galois}
\end{align*}
\begin{align*}
& a \sas b \leq a \sas b \\
\Rightarrow \ & (a \sas b)^\bot \sas b \leq a^\bot & \text{Galois} \\
\Rightarrow \ & ((a \sas b)^\bot \sas b) \sas b \leq a^\bot & \text{Orthomodularity} \\
\Rightarrow \ & (((a \sas b)^\bot \sas b) \sas b) \sas b \leq a^\bot & \text{Orthomodularity} \\
\Rightarrow \ & a \sas b \leq (((a \sas b)^\bot \sas b) \sas b)^\bot & \text{Galois} \\
\Rightarrow \ & a \sas b \leq (((a \sas b)^\bot \sas b) \sas b)^\bot \sas b & \text{Orthomodularity}
\end{align*}
\end{Proof}

This characterization of orthomodular lattices in terms of Sasaki orthoposets shows that this notion can entirely be characterized using only elements taken from classical but partial observation of the world, the main hypothesis being that there are ``enough'' points of view. In particular, this constitutes a way to envision orthomodularity in a purely classical manner, where the key ideas are the notions of points of view and of partiality of knowledge.

In the following, we will use the structure of Sasaki orthoposets to introduce what we call the Sasaki orthologic, first defined in a classical axiomatic way, and then in a rule-based manner.

\section{The Sasaki Orthologic}

\subsection{Basic Definitions}

Let us first provide some usual definitions. Given a set $\Psi$ of atomic propositions, we define the language $\mc L_\Psi$ as the collection of terms defined using the following grammar:
$$ t = a \,|\, t \, \& \, t \,|\, t^\bot $$
where $a$ is an element of $\Psi$. Now, we define a Sasaki model as a pair formed by a Sasaki orthoposet, and an assignment of atomic propositions onto this Sasaki orthoposet:

\begin{Def}[Sasaki Model]
A $\Psi$-Sasaki model is a pair $\mc M = \pair{\tuple{P,\leq_P,\bot_P,\&_P}} \nu$ where $\tuple{P,\leq_P,\bot_P,\&_P}$ is a Sasaki orthoposet and $\nu: \Psi \rightarrow P$ is a function mapping atomic propositions to elements of the Sasaki orthoposet.
\end{Def}

We also define $\text{SM}(\Psi)$ as the set of $\Psi$-Sasaki models.

\begin{Def}[Interpretation Function]
Given a $\Psi$-Sasaki model $\mc M = \pair {\tuple{P,\leq_P,\bot_P,\&_P}} \nu$, we define the interpretation function $\sem \cdot_{\mc M}: \mc L_{\Psi} \rightarrow P $ inductively as:
\begin{align*}
\fall{a \in \Psi} \sem a_{\mc M} & = \nu(a) \\
\sem {t^\bot}_{\mc M} & = \paren{\sem t_{\mc M}}^{\bot_P} \\
\sem{t_1 \,\& \, t_2}_{\mc M} & = \sem {t_1}_{\mc M} \,\&_P\, \sem{t_2}_{\mc M}
\end{align*}
\end{Def}

\begin{Def}[Validity]
An inequality $t_1 \leq t_2$ is valid with regards to Sasaki Orthologic (which we denote $ \vdash_{\rm SOL} t_1 \leq t_2$) if and only if:
$$ \paren{\fall {\mc M \in \text{SM}(\Psi)} \sem{t_1}_{\mc M} \leq_{\mc M} \sem{t_2}_{\mc M}} $$
where given a Sasaki model $\mc M$, we identify the relation $\leq_{\mc M}$ with the partial order relation of its underlying Sasaki orthoposet.
\end{Def}

\subsection{A Rule-Based Definition of Sasaki Orthoposets}

We now turn to the definition of a rule-based logic for characterizing Sasaki orthoposets.

\begin{figure}
\begin{gather*}
\prooftree \vphantom{a^\bot} \justifies a \leq a \using A \endprooftree \quad \quad
\prooftree b^\bot \leq a^\bot \justifies a \leq b \using S \endprooftree \quad \quad
\prooftree c^\bot \sas b \leq a^\bot \justifies a \sas b \leq c \using G \endprooftree \\
\ \\
\prooftree a \leq b \justifies a^{\bot\bot} \leq b \using N_L \endprooftree \quad \quad
\prooftree a \leq b \justifies a \leq b^{\bot\bot} \using N_R \endprooftree \quad \quad
\prooftree a \leq b \quad b \leq c \justifies a \leq c \using T \endprooftree \\
\ \\
\prooftree a \leq b \quad a \leq c \justifies a \sas b \leq c \using O_L \endprooftree \quad \quad
\prooftree a \leq b \quad a \leq c \justifies a \leq b \sas c \using O_R \endprooftree \\
\ \\
\prooftree b \leq c \justifies a \sas b \leq c \using R \endprooftree \quad \quad
\prooftree a \leq c \quad b \leq d \quad d \leq b \justifies a \sas b \leq c \sas d \using M \endprooftree
\end{gather*}
\caption{Rules for {\em RSOL}} \label{Fig:Rules}
\end{figure}

\begin{Def}[RSOL]
Let $\rm RSOL$ (for {\em Rule-based Sasaki Orthologic}) be the logic defined by the rules given in figure \ref{Fig:Rules}. In other words, given two terms $a,b \in \mc L_\Psi$, the inequality $a \leq b$ is valid in $\rm RSOL$ (which we denote $ \vdash_{\rm RSOL} a \leq b $) if and only if this inequality can be proved using the rules given in figure \ref{Fig:Rules}.
\end{Def}

\begin{Prop}[Soundness]
$\rm RSOL$ is sound w.r.t. $\rm SOL$, that is for all $a,b \in \mc L_\Psi$, one has:
$$ \vdash_{\rm RSOL} a \leq b \ \Rightarrow \,\ \vdash_{\rm SOL} a \leq b $$
\end{Prop}
\begin{Proof}
This follows from the fact that every rule in figure \ref{Fig:Rules} is valid in $\rm SOL$. More precisely, rules $A$, $S$, $N_L$, $N_R$ and $T$ come from the definition of an orthoposet. Rule $G$ derives from the {\em Galois} property, $O_L$ and $O_R$ from the {\em Orthomodularity} property, $R$ from the {\em R-Reduction} property and finally $M$ derives from the {\em L-Monotony} property.
\end{Proof}

In order to prove that it is also complete w.r.t. $\rm SOL$, we introduce a few notations and definitions. First, given a term $t \in \mc L_\Psi$, we define the set $\sem t$ of terms equivalent to $t$ w.r.t. to {\rm RSOL}:
$$\sem t = \set {u \in \mc L_\Psi}{\vdash_{\rm RSOL} t \leq u \cand \vdash_{\rm RSOL} u \leq t} $$
Moreover, let us define:
$$ {\mc L_\Psi}_{/ R} = \set {\sem t}{t \in \mc L_\Psi} $$
$$ \fall {a,b \in \mc L_\Psi} \sem a \leq_{/ R} \sem b \ \Leftrightarrow \,\ \vdash_{\rm RSOL} a \leq b $$
$$ \fall {a \in \mc L_\Psi} {\sem a}^{\bot_{/ R}} = \sem {a^\bot} $$
$$ \fall {a,b \in \mc L_\Psi} {\sem a} \, \&_{/ R} \, {\sem b} = \sem {a \sas b} $$
Because of rules $S$ and $M$, the definition of $\bot_{/ R}$ and $\&_{/ R}$ make sense.
\begin{Prop} \label{Prop:CanonicSOP}
The tuple $\tuple{{\mc L_\Psi}_{/R}, \leq_{/R}, \bot_{/R}, \&_{/R}}$ is a Sasaki orthoposet.
\end{Prop}
\begin{Proof}
It is clear that $\tuple{{\mc L_\Psi}_{/R}, \leq_{/R}, \bot_{/R}}$ is an orthoposet. The following prooftrees show that the four properties concerning $\&$ hold in the present situation:
$$
\begin{tabular}{cl}
\prooftree a \leq b \quad \prooftree \justifies c \leq c \using A \endprooftree \quad \prooftree \justifies c \leq c \using A \endprooftree \justifies a \sas c \leq b \sas c \using M \endprooftree & \text{\em L-Monotony} \bigskip \\
\prooftree \prooftree \justifies b \leq b \using A \endprooftree \justifies a \sas b \leq b \using R \endprooftree & \text{\em R-Reduction} \bigskip \\
\prooftree a \leq b \quad \prooftree \justifies a \leq a \using A \endprooftree \justifies a \sas b \leq a \using O_L \endprooftree \quad \quad \prooftree \prooftree \justifies a \leq a \using A \endprooftree \quad a \leq b \justifies a \leq a \sas b \using O_R \endprooftree & \text{\em Orthomodularity} \bigskip \\
\prooftree a \sas b \leq c \justifies c^\bot \sas b \leq a^\bot \using G \endprooftree & \text{\em Galois}
\end{tabular}
$$
\end{Proof}

\begin{Prop}[Completeness]
$\rm RSOL$ is complete w.r.t. $\rm SOL$, that is for all $t,u \in \mc L_\Psi$, one has:
$$ \vdash_{\rm SOL} t \leq u \ \Rightarrow \,\ \vdash_{\rm RSOL} t \leq u $$
\end{Prop}
\begin{Proof}
This is a consequence of proposition \ref{Prop:CanonicSOP}, since $\tuple{{\mc L_\Psi}_{/R}, \leq_{/R}, \bot_{/R}, \&_{/R}}$ can be turned into a $\Psi$-Sasaki model by using the mapping $a \mapsto \sem a$ for atomic propositions.
\end{Proof}

Thus, we have seen that $\rm RSOL$ constitutes a sound and complete axiomatization of Sasaki orthoposets or equivalently, of orthomodular lattices. It is not yet known whether this logic is decidable or not. However, from its definition, it is possible to exhibit a decidable fragment. as we now show.

\subsection{A Decidable Fragment of $\rm RSOL$}

In the definition of $\rm RSOL$, the $T$ rule plays a special role since it is the only one in which the premisses contain terms which are not in the conclusion. This suggest to define a fragment of this logic which omits the $T$ rule.

\begin{Def}[${\rm RSOL}/T$]
Let ${\rm RSOL}/T$ (for $\rm RSOL$ minus the $T$-rule) be the logic defined by the rules $A$, $S$, $G$, $N_L$, $N_R$, $O_L$, $O_R$, $R$ and $M$ given in figure \ref{Fig:Rules}.
\end{Def}

\begin{Prop}
${\rm RSOL}/T$ is decidable.
\end{Prop}
\begin{Proof}
This follows from the fact that every proof in ${\rm RSOL}/T$ is of finite height, since:
\begin{itemize}
\item The logical rules ($O_L$, $O_R$, $R$ and $M$) do all verify the subformula property and their premisses contain strictly less occurences of the $\&$ connective than their conclusion.
\item The structural rules ($S$, $G$, $N_L$ and $N_R$) are involutive, so that one can suppose that the same structural rule does not occur twice in a row.
\item Regardless of $N_L$ and $N_R$, given a proof whose last rules are made of $n$ successions of $S$ and then $G$, its conclusion has to be of the form $(\ldots (a \sas b_1) \sas b_2 ) \ldots \sas b_n \leq c$. As a consequence, there cannot be infinitely many structural rules at the end of a proof.  
\end{itemize} 
As a consequence, in a ${\rm RSOL}/T$ proof, there are finitely many occurences of logical rules and there are finitely many occurences of structural rules between any two successive logical rules.
\end{Proof}

\begin{Prop} \label{Prop:TAdmiss}
If the $T$-rule from figure \ref{Fig:Rules} is admissible in ${\rm RSOL}/T$, then ${\rm RSOL}$ is decidable.
\end{Prop}

The $T$-rule, which has been omitted in the definition of ${\rm RSOL}/T$, corresponds to the {\em Cut} rule in usual logical systems. As such, proposition \ref{Prop:TAdmiss} shows that it would be extremely interesting to study the possible elimination of cuts in this logic.

\section{Conclusion and perspectives}

In this article, we have presented a way to characterize orthomodular lattices by focusing on the binary operation called the {\em Sasaki projection} rather than using usual lattice operations such that the meet and the join.

After having given an abstract definition of the {\em Sasaki projection}, we have introduced a structure called {\em Sasaki orthoposet} and have shown that they are equivalent to orthomodular lattices. Then, we have introduced a rule-based @@@

This work provides a new direction in the study of the logic corresponding to orthomodular lattices (which is usually called standard quantum logic). The rule-based formalism, which has many similitudes with Gentzen's sequent calculus for more general logics, permits to explore its decidability (as suggested by the question of the possibility of cut elimination) and can serve as a basis for implementing automated proof checkers for this logic.

Moreover, one can consider several extensions to this formalism (with for instance statements of the form {\em ``if an orthomodular lattice verifies inequalities $t_1 \leq u_1$, ..., $t_n \leq u_n$, does it verify $t \leq u$?''}) and use it to explore notions like the tensorial product of orthomodular lattices, or the difference between orthomodular and Hilbert lattices.

\bibliographystyle{apalike}

\end{document}